\documentclass[12pt]{iopart}

\usepackage{graphicx}
\usepackage{iopams}
\begin{document}

\title[Non analytical $f(R)$ theory of gravity]{Non-analytical power law correction\\ to the Einstein-Hilbert action:\\ gravitational wave propagation}

\author{Donatella Fiorucci$^{1a}$, Orchidea Maria Lecian$^{2b}$ and Giovanni Montani$^{134c}$ }

\address{
$^{1}$ Dipartimento di Fisica Universit\`a di Roma ``Sapienza'',\\ Piazzale Aldo Moro 5, 00185, Roma, Italy}
\address{$^{2}$ Max Planck Institute for Gravitational Physics (Albert Einstein Institute),\\ am M\"uhlenberg 1, D-14476 Golm, Germany}
\address{$^{3}$ ENEA - C.R. Frascati, U.T.Fus. (FUSMAG Lab),\\ Via Enrico Fermi, 45 (00044), Frascati (Roma), Italy}
\address{$^{4}$ INFN - Istituto Nazionale di Fisica Nucleare, Sezione di Roma 1}
\ead{\mailto{$^{a}$donatellafiorucci@ymail.com}}
\ead{\mailto{$^{b}$omlecian@aei.mpg.de}}
\ead{\mailto{$^{c}$giovanni.montani@frascati.enea.it}}

\begin{abstract}
We analyze the features of the Minkowskian limit of a particular non-analytical
$f(R)$ model, whose Taylor expansion in the weak field limit does not hold, as far as gravitational waves (GWs) are concerned. We solve
the corresponding Einstein equations and we find an explicit expression
of the modified GWs as the sum of two terms, i.e. the standard one and
a modified part.
As a result, GWs in this model
are not transverse, and
their polarization is different from that of General Relativity.
The velocity of the GW modified part depends crucially on the parameters characterizing the model, and it mostly results much smaller than the speed of light.
Moreover, this investigation allows one to further
test the viability of this particular $f(R)$ gravity theory as far as
interferometric observations of GWs are concerned.
\end{abstract}


\maketitle

\section{Introduction}
Within the possible modifications of General Relativity (GR), $f(\mathbf{R})$ modified theories of gravity are based on replacing the Ricci scalar $\mathbf{R}$ in the Einstein-Hilbert (EH) action by a suitable function $f(\mathbf{R})$ of it. 
Even though the predictions of GR reproduce extremely successfully the great majority of the phenomena within a very stringent accuracy, there are nevertheless some experimental data which are not matched by GR. Some of these phenomena can be explained as due to the presence of extra (otherwise unobserved) matter or energy contributions. $f(\mathbf{R})$ theories of gravity are a tool to describe these phenomena as due to effects of the geometry (i.e. 'on the left-hand side of Einstein equations') rather than of some kind of matter (i.e. 'on the right-hand side of Einstein equations'). There are at the moment many analyses aimed at discriminating from an experimental point of view between geometrical effects and matter effects.\\
It is then important to constrain the parameter space of these models in such a way to reproduce these unexpected phenomena without discarding all the other well-tested predictions of GR. Moreover, it is important to stress that the suitably-constrained parameter space should be valid at all physical scales.\\
It is interesting to remark that some kind of $f(\mathbf{R})$ theories of gravity descend from a suitable 'low energy limit' of some other unification theories. We remark that, in the latter case, the presence of other invariants (built as suitable products of the Riemann and the Ricci tensor as well) is also possible \cite{Nojiri:2006je}.\\  
In this paper, we study the behaviour of the
gravitational waves in the non-analytical $f(\bf{R})$
model proposed in \cite{Lecian:2008vc}, i.e. $f(\mathbf{R})=\mathbf{R}+\gamma \mathbf{R}^{\beta}$, since
the corresponding weak field equations posses a
peculiar feature, for which retaining non-linear
terms in the dynamics makes sense.
In fact, as far as the parameter $2 < \beta < 3$,
there are non-linear corrections to the General Relativity
analysis which, being greater than quadratic
term and tending to the same order of the linear ones
as $\beta \rightarrow 2$, can not be disregarded
and have a significant impact on the theory predictions
in the limit of weak spacetime ripples propagation.

Our study outlines how, in parallel with the linear
vacuum massless gravitational waves of General Relativity, a non-linear
wave appears, having a non-trivial
(non-transverse and non-traceless) morphology and whose
amplitude increases with time.
The effect induced by this ``anomalous'' wave
on the test particle geodesic deviation is
described in some detail, and the request that the
modified tidal term be below the typical experimental
constraints lead us to restrict the velocity
range of these non-linear ripples.
Indeed, a peculiarity of the considered model
consists of the non-null character of the propagation
vector, whose modulus results to be fixed by
the initial and boundary conditions on the ``radiation'' field,
and must take values corresponding to a propagation velocity well-below the speed of the light,
apart from the limiting case $\beta \rightarrow 2$.

Finally, we provide a study of the polarization
of the modified wave, based on the action of the
tidal force on a particle system initially arranged on a circumference and
having as reference the comparison to
the standard case. This characterization of the
anomalous deformation of the polarization ellipses
offers a valuable tool for recognizing the
presence of this non-linear feature among the
background signals detected by ground interferometers.
In particular, the increasing power-law behaviour
that the modified contribution outlines at a fixed
point of space makes this new effect, whether existing,
very promising for a future detection. In this
sense, the present study, on the one hand, offers a very
reliable trace to identify a specific gravitational
dynamics modification, and, on the other hand, offers an intriguing scenario for enhancing the detection
of gravitational signals, which, in the
considered case, are particularly able to bring
information on their generating source,
here schematized by the initial and boundary condition
on their propagation.

The paper is organized as follows.
In Section \ref{section2}, we briefly recall the basic statements about $f(\mathbf{R})$ theories of gravity, with particular attention to their weak-field limit. In Section \ref{section3}, we review the main features of a particular class of non-analytical $f(\mathbf{R})$ theories of gravity, i.e. $f(\mathbf{R})=\mathbf{R}+\gamma \mathbf{R}^\beta$, where the standard EH action of GR is perturbed by the addition of a term, for which the Taylor series in the vicinity of $R=0$ does not hold. In section \ref{section4}, we first briefly review the main features of the weak-field limit of $f(\mathbf{R})$ models, analytically solve and discuss the weak-field limit of the Einstein equations for this model, and explicitly write down the form of gravitational waves; to do so, we analyze the physical meaning of the degrees of freedom of the model. We also compare our results with those obtained in the case of analytical $f(\mathbf{R})$ models. In Section \ref{section5}, we discuss the interaction of gravitational waves with test particles. Brief concluding remarks end the paper.   

\section{$f(\textbf{R})$ modified theories of gravity\label{section2}}
$f(\bf{R})$ theories of gravity are obtained when the Ricci scalar $\textbf{R}$ in the Einstein-Hilbert action
\begin{equation}
S_{EH}= -\frac{c^3}{16 \pi G}\int d^4x \sqrt{-\bf{g}} \bf{R}  ,
\end{equation}
is replaced by an arbitrary function $f(\bf{R})$ of it, such that the modified gravitational action reads
\begin{equation}
\label{eq:mfe}
S={-\frac{c^3}{16\pi G}}\int d^{4}x\sqrt{-\bf{g}}f(\bf{R}),
\end{equation}
where $\sqrt{-\bf{g}}$ is the determinant of the metric tensor $\bf{g}_{\mu\nu}$, $\sqrt{-\bf{g}}\equiv det \bf{g}_{\mu\nu}$. In vacuum, the modified Einstein equations take the form
\begin{equation}
\label{gfe}
\eqalign{
&{-\frac{1}{2}}\mathbf{g}_{\mu\nu}f(\mathbf{R})+f'(\mathbf{R})\mathbf{R}_{\mu\nu}-\nabla_{\mu}\nabla_{\nu}f'(\mathbf{R})+\mathbf{g}_{\mu\nu}\mathbf{g}^{\rho\sigma} \nabla_{\rho} \nabla_{\sigma} f'(\mathbf{R}) = 0,\\
&3\mathbf{g}^{\rho\sigma} \nabla_{\rho} \nabla_{\sigma}f'(\mathbf{R})+f'(\mathbf{R})\mathbf{R}-2f(\mathbf{R}) = 0,}
\end{equation}
where $\mathbf{R}_{\mu\nu}$ is the Ricci tensor, $\nabla_{\mu}$ is the covariant derivative, and a prime denotes differentiation with respect to $\mathbf{R}$, $f'(\mathbf{R})=df/d\mathbf{R}$. Here and in the following, we will use bold letters for the full covariant objects (such as $\mathbf{g}_{\mu\nu}$, $\mathbf{R}_{\mu\nu}$, $\mathbf{R}$), and usual letters for their weak-field expression (such as $g_{\mu\nu}$, $R$, $R_{\mu\nu}$). We will adopt the following notation: Greek indices run form $0$ to $3$, i.e. $\mu,\nu=0,1,2,3$, Latin indices run form $1$ to $3$, i.e. $i,j=1,2,3$, and we follow the standard notation \cite{Landau}, with mostly-minus signature.

The physical effect of $f(R)$ models is to add a scalar degree of freedom to EH gravity (see, for example, the review \cite{Sotiriou:2008rp} and the references therein). In the Jordan frame, the dynamical features of the non-constant first derivative $f'(R)$ can be treated as a scalar field. In the scalar-tensor version of the models, via a conformal transformation, the modified action (\ref{eq:mfe}) is rewritten in terms of a scalar field minimally coupled to gravity in the Einstein frame.

\paragraph{General weak-field limit}
In the weak-field limit, in vacuum, we consider the metric tensor $g_{\mu\nu}$ as consisting of the flat metric $\eta_{\mu\nu}=(1,-1,-1,-1)$ and a small perturbation of it, $|{h}_{\mu\nu}|<<1$, i.e. 
\begin{equation}
\label{eq:1a}
g_{\mu\nu}={\eta}_{\mu\nu}+{h}_{\mu\nu}.
\end{equation}
The weak-field limit of Einstein equations consists in considering the $\mathcal{O}(h)$-terms only, and in neglecting higher-order $\mathcal{O}(h^2)$-terms.

If the function $f(\mathbf{R})$ admits a Taylor expansion in the neighborhood of $R\sim0$, its Taylor series $f(R)=\sum_{j=0}^{j=\infty}a_j R^j$ (where the $a_j$'s are the $j$-th order Taylor coefficients of the series and have the dimension of {\it length} $^{2j-2}$) can be inserted in the modified Einstein equations.\\
Even though an infinite number of parameters has, in principle, to be fixed, the weak-field limit of this class of $f(\mathbf{R})$ models can be shown to depend only on the parameters $a_1$ and $a_2$, for vanishing $a_0$. In fact, the limit of flat Minkowski space (in GR) is recovered for $R\sim0$, such that $a_0\sim0$.\\
Furthermore, we remark that, even though any value of the parameter $a_1$ is in principle admitted from the Taylor expansion, if the Taylor series of $f(R)$ has to be interpreted as a perturbation of the Ricci scalar in the standard EH action, then the value of $a_1$ should be close to $1$. In fact, the effect of $a_1\neq1$ can be interpreted, in the comparison with GR, as a modification of the value of the gravitational constant $G$. (See \cite{Teyssandier}, \cite{Stelle}). 

If, on the contrary, the function $f(\mathbf{R})$ is not analytical around $R\sim0$ (i.e. its Taylor expansion does not hold in the neighborhood of $R\sim0$), other paradigms are to be looked for. It is within this perspective that non-analytical $f(\mathbf{R})$ models have been addressed.\\
For example, the model $f(\mathbf{R}) = \mathbf{R}-\mu^2\sin\frac{\mu^2}{\mathbf{R}-\Lambda}$ (where $\mu$ is a constant with the dimensions of \textit{length}$^{-1}$ and $\Lambda$ is a constant with the dimensions of \textit{length}$^{-2}$) has been proposed in \cite{Jin:2006if}.\\
Furthermore, the $f(\mathbf{R})$ model consisting of a sum of different powers of the Ricci scalar, such as $f(\mathbf{R})=\mathbf{R}+a_n\mathbf{R}^{n}+a_m\mathbf{R}^{m}$ (where the constants $a_j$ have the dimensions of \textit{length}$^{2j-2}$) has been investigated in \cite{Srivastava:2007bc}, \cite{Fay:2007gg} and \cite{Nojiri:2003ft} as far as cosmological implications are concerned. No analytical solution for the weak-field limit of the corresponding field equations exists, not even for (physically-relevant) special choices of the exponents $n$ and $m$.  

\section{Non-analytical $f(\bf{R})$ models\label{section3}}
In this section, we briefly review the main features of an example of non-analytical $f(\bf{R})$ model, namely
\begin{equation}\label{rbeta}
f(\bf{R})=\bf{R}+\gamma\bf{R}^\beta,
\end{equation}
where the parameter $\gamma$ has the dimension of {\it length} $^{2\beta-2}$, and the exponent $\beta$ is dimensionless. The typical \textit{length}-scale $L_\gamma$ of the model can be worked out of the $\gamma$ parameter as $L_\gamma\equiv\mid\gamma\mid^{\frac{1}{2\beta-2}}$. The interest in this model is based on the fact that field equations are analytically solvable, as in \cite{Lecian:2008vc} and \cite{Capozziello:2011yr}.

The weak-field limit of (\ref{rbeta}) was studied in \cite{Lecian:2008vc}. The weak-field limit of field equations
can be solved consistently as a perturbation of the flat metric only for $2<\beta<3$. In fact, for $\beta\ge3$, the correction would be $\ge\mathcal{O}(h^2)$, for $\beta<2$ the correction would be grater than the ordinary $\mathcal{O}(h)$ terms for $R\sim0$, while for $\beta=2$ the analytical case would be recovered.

Because of the properties of the modified Einstein equations and of the functional dependence of (\ref{rbeta}) on $\bf{R}$, it is straightforward to understand that the solution of the pertinent Einstein equations in the weak-field limit will consist of two parts, namely the standard GR term plus a correction term. In particular, in \cite{Lecian:2008vc}, the spherically-symmetric Einstein equations were solved, and the generalized gravitational potentials were found to consist of a Schwarzschild term (responsible for the Newtonian behaviour) plus a modification term, whose features depend crucially on the $\gamma R^\beta$ term. The parameter space of the model was constrained by
imposing compatibility with Solar-System data. As a result, a wide range for the values of the parameter $\gamma$, hence of the characteristic length scale $L_{\gamma}$, was demonstrated to exist, and a lower bound was determined.

\begin{figure}[ht!]
\centering
\includegraphics[width=0.78\textwidth, height=0.33\textheight]{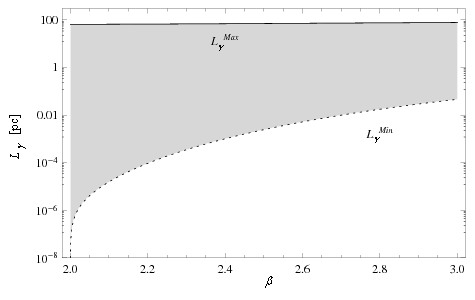}
\caption{\label{fig:lgamma}\small{The gray zone represents the allowed characteristic-length scales of the model $f(\bf R)=\bf R+\gamma \bf R^{\beta}$. $L^{Min}_{\gamma}$ comes from the Solar-System data and $L^{Max}_{\gamma}$ comes from the study of the cosmological implications. See \cite{Capozziello:2011yr}.}}
\end{figure}

The cosmological implementation of this same model has been addressed in \cite{Capozziello:2011yr}. In particular, the possibility to recover both a radiation-dominated era and a matter-dominated one has been demonstrated. From these calculations, also an upper bound for $L_\gamma$ has been found. Combining the two results, it is possible to determine the allowed values for the characteristic length scale $L_{\gamma}$ as a function of $\beta$ (see figure (\ref{fig:lgamma}), which was obtained in \cite{Capozziello:2011yr}).
Furthermore, the gravitational-wave evolution on a RW background has been illustrated to be, for all practical purposes, the same as in GR, and any contribution to the inflationary GW background was shown to stay below the detection threshold of present and future interferometers.

In the following, we will analyze the weak-field limit of the model as far as the presence of gravitational waves is concerned.  

\section{Minkowskian limit\label{section4}}
We will now solve the field equations in the weak field limit, and describe the main features of gravitational waves. To do so, we will recall the fundamental gauges of GR, which is possible to impose in order to get physical insight on the solution of field equations. After this, we will briefly recall the known results for analytical models, for which a comparison is useful. We will then find the solutions of the field equations in the non-analytical case, and discuss them both on the basis of their physical interpretation and of their mathematical well-posed-ness.\\
In the weak-field limit, we can evaluate the Ricci tensor, $R_{\mu\nu}$, and the Ricci scalar, $R$, by retaining terms that are first order in ${h}_{\mu\nu}$ only. Thus, we get the expressions of the Ricci tensor $R_{\mu\nu}$ and of the Ricci scalar as a function of the perturbation $h_{\mu\nu}$, respectively:
\begin{equation}
\label{wflim}
\eqalign{
&R_{\mu\nu}=\frac{1}{2}\left[-\Box h_{\mu\nu}+\left({h^\alpha}_{\nu, \mu\alpha}+{h^\alpha}_{\mu, \nu\alpha}-{h^\alpha}_{\alpha, \mu\nu}\right)\right],\\
&R=-\Box h+{h^{\alpha\nu}}_{,\alpha\nu}\ ,}
\end{equation}
where $\Box\equiv\eta^{\mu\nu}\partial_\mu\partial_\nu$ is the flat-space D'Alembertian operator \footnote{In the present paper, the definition of the flat-space D'Alembertian operator differs from that given in \cite{Landau}, where $\Box\equiv -\eta^{\mu\nu}\partial_\mu\partial_\nu$.}.

The Einstein equations (\ref{gfe}) for our model (\ref{rbeta}) read:
\begin{equation}
\label{ees}
\eqalign{
&\bf{R}_{\mu\nu}-\frac{1}{2}\bf{g}_{\mu\nu}\bf{R}-\gamma\beta{\nabla_{\mu}}{\nabla _{\nu}}\bf{R}^{\beta-1}+\gamma\beta \bf{g}_{\mu\nu} \bf{g}^{\rho\sigma}{\nabla_{\rho}}{\nabla _{\sigma}}\bf{R}^{\beta-1}=0,\\
&3\gamma\beta \bf{g}^{\mu\nu}{\nabla_{\mu}}{\nabla _{\nu}}\bf{R}^{\beta-1}-\bf{R}+\gamma(\beta-2) \bf{R}^{\beta},}
\end{equation}
and, in the weak-field limit\footnote{We remark that the weak field limit of the differential operator $g^{\mu\nu}\nabla_\mu\nabla_\nu$ (which appears in (\ref{ees}) acting on (the weak-field limit of) an object $X$ such that $\mathcal{O}(h)\le X<\mathcal{O}(h^2)$ is the flat-space D'Alembertian operator $\Box\equiv\eta^{\mu\nu}\partial_\mu\partial_\nu$ (as in (\ref{ees2})-(\ref{ees3})).}, they become:
\numparts
\begin{equation}
R_{\mu\nu}-\frac{1}{2}\eta_{\mu\nu}R-\gamma\beta{\nabla_{\mu}}{\nabla _{\nu}}R^{\beta-1}+\eta_{\mu\nu}\gamma\beta \Box R^{\beta-1}=0, \label{ees2}
\end{equation}
\begin{equation}
3\gamma\beta \Box R^{\beta-1}=R, \label{ees3}
\end{equation}
\endnumparts
This way, it is possible to rewrite Einstein equations (\ref{ees2})-(\ref{ees3}) for the physical unknowns $h_{\mu\nu}$ and $h$ through he weak-field expansion (\ref{wflim}).

\paragraph{Analysis of the allowed gauges}
The features of the physical unknowns $h_{\mu\nu}$ is at the basis of the determination of the possible gauges that can be imposed.\\
We recall that the 'de Donder' gauge allows one to reduce the expression of the Ricci tensor $R_{\mu\nu}$ and that of the Ricci scalar $R$ to
\begin{equation}
\label{wfl}
\eqalign{
&R_{\mu\nu}=-\frac{1}{2} \Box h_{\mu\nu},\\
&R=-\frac{1}{2}\Box h
}
\end{equation}
where $\Box$ is the flat-space D'Alembertian operator, by imposing $g^{\mu\nu}{\Gamma^{\alpha}}_{\mu\nu}=0$, i.e. $h^\mu_{\ \ \nu,\mu}=\frac{1}{2}h^\lambda_{\ \ \lambda,\nu}$. Because of the specific expression of field equations in General Relativity, this gauge is also known as the 'harmonic gauge' \textit{because} Einstein equations in vacuum are solved by harmonic functions for $h_{\mu\nu}$.

 As well-known, in modified theories of gravity, the solution of field equations is not given \textit{only} by harmonic functions. Nevertheless, because the de Donder gauge is based on the discussion of the degrees of freedom, we can apply the same reasoning to the modified field equations.\\
Because of the linearity properties of the D'Alembertian operator $\Box$, it is always possible to split the perturbations $h_{\mu\nu}$ into two parts, i.e.
\begin{equation}\label{harm}
h_{\mu\nu}\equiv h^{(0)}_{\mu\nu}+h^{(1)}_{\mu\nu},
\end{equation}
where $h^{(0)}_{\mu\nu}$ is a harmonic function, i.e. $\Box  h^{(0)}_{\mu\nu}\equiv0$, while $h^{(1)}_{\mu\nu}$ is not, i.e. $\Box  h^{(1)}_{\mu\nu}\neq0$. An analogous discussion can be made for $h\equiv\eta^{\mu\nu}h_{\mu\nu}\equiv h^{(0)}+h^{(1)}$.

We remark that the non-trivial part of field equations (\ref{ees2})-(\ref{ees3}) contain $h_{\mu\nu}$ and $h$ only through the expansion (\ref{wfl}), i.e. they account only for the non-harmonic functions $h^{(1)}_{\mu\nu}$ and $h^{(1)}$.\\
Furthermore, the sum of two harmonic functions is still a harmonic function, by which the harmonic part $h^{(0)}_{\mu\nu}$ can be suitably defined. Accordingly, because of the explicit expression of the field equations (\ref{ees2})-(\ref{ees3}) as functions of $h_{\mu\nu}$ and $h$ through (\ref{wfl}), it is physically equivalent (apart from the order of the corresponding equations) to discuss the field equations for the physical unknowns $h_{\mu\nu}$ and $h$ or for the geometrical objects $R_{\mu\nu}$ and $R$, defined as in (\ref{wfl}).

The 'transverse-traceless' (TT) gauge can be imposed on $h^{(0)}_{\mu\nu}$ by using the four degrees of freedom left by the infinitesimal transformation of the coordinates, which is performed when fixing the de Donder gauge. Thus we get $h^{(0)TT}_{\mu\nu}$.

\paragraph{Analytical case}
We briefly discuss here the main results of the Minkowskian limit of analytical $f(R)$ models.\\
The weak-field limit of field equations for the unknowns $R_{\mu\nu}$ and $R$ read
\begin{equation}
\eqalign{
&R_{\mu\nu}=\frac{1}{6}\eta_{\mu\nu}R+2a_2R,_{\mu\nu},\\
&\Box R=\frac{a_1}{6a_2}R,
}
\end{equation}
which are formally solved for the physical unknowns $h^{(1)}_{\mu\nu}$ (once the solution for $R$ is taken into account) as
\numparts
\begin{equation}
h_{\mu\nu}^{(1)}=-\frac{2}{q_0^2-q^2}\left(\frac{1}{6}\eta_{\mu\nu}R+2a_2q_\mu q_\nu R\right),\label{hmn1c}
\end{equation}
\begin{equation}
R=Ae^{iq^\mu x_\mu}\label{r2},
\end{equation}
\endnumparts
$A$ being an integration constant with the dimensions of \textit{length}$^{-2}$.\\
The solutions (\ref{hmn1c}) and (\ref{r2}) of field equations imply the presence of a harmonic part $h^{(0)}_{\mu\nu}$ and a modified part $h^{(1)}_{\mu\nu}$ consisting of a  massive mode $e^{iq_\mu x^\mu}$ with mass $m$, such that $q_0^2-q^2\equiv m^2$, with several possible polarizations. It is interesting to remark, for future purposes, that the solution
(\ref{hmn1c}) consists of two different parts: one containing the
flat-space tensor $\eta_{\mu\nu}$, and one containing the object
$q_\mu q_\nu$. In particular, the latter is interpreted as a
(constant) polarization tensor. Furthermore, this is obtained, in the
solution of field equations, from the derivatives of the Ricci scalar
(\ref{r2}). The functional dependence of the Ricci scalar (\ref{r2})
on $x^\mu$ such that $R\equiv R(q^\mu x_\mu)$ is the only one that
allows one to obtain a constant polarization tensor $q_\mu q_\nu$ from
the derivatives of $R$.\\
Despite solution (\ref{hmn1c}) is obtained directly from field equations, in the literature different assumptions have sometimes been made, such as those in \cite{Capozziello:2008rq} and \cite{Berry:2011pb}. A mathematical classification of the possible (extra) polarizations that can appear in the weak-field limit of $f(R)$ models, according to the features of the solution of the geometrical objects $R_{\mu\nu}$ and $R$, has been performed in \cite{Alves:2009eg}.
The behaviour of Weyl scalars in GR in the case of a binary black hole
inspiral and merger has recently been numerically investigated in \cite{eloisa}.
The comparison with the case of $f(R)$ models could be a fruitful tool
to further test the viability of modified theories of gravity in these
scenarios.

It is straightforward to remark that the scalar degree of freedom introduced by $f(R)$ models allows one to rewrite the trace equation for the scalar degree of freedom, say $\Psi$, as
\begin{equation}
\Box \Psi=m^2\Psi
\end{equation}
where one identifies $\Psi=R$, and the term $m^2\Psi$, with $m^2\equiv a_1/(6a_2)$ the effective mass of the scalar field, as the linearization of a potential $V(\Psi)$ (ruling the dynamics of $\Psi$) around its minimum, which corresponds to $R\sim0$. The fact that an $f(R)$ is analytical, i.e. its Taylor expansion holds in the neighborhood of $R\sim0$, is therefore equivalent to linearizing the dynamics of the scalar degree of freedom $\Psi$ on a fixed background, where the potential $V(\Psi)$ \textit{does admit} a Taylor expansion in the neighborhood of such a minimum.\\
This way, the choice of (\ref{r2}) accounts for a term describing a propagating massive field in (\ref{hmn1c}).

\subsection{Non-analytical weak field limit}  
We can now solve the weak-field limit of the Einstein equations for our model, (\ref{ees2}-\ref{ees3}).\\
The trace equation (\ref{ees3}) is solved by
\begin{equation}\label{tracer}
R=\left(\delta\xi\right)^{\frac{2}{\beta-2}},
\end{equation}
where the variable $\xi$ has the dimension of a \textit{length}, the constant $\delta$ has the dimension of \textit{length}$^{1-\beta}$, and read
\begin{equation}
\eqalign{
\label{eq:xi}
&\xi=(C_2+q_{\mu}x^{\mu}),\\
&\delta=\left(\frac{1}{6\gamma{(}q_{0}^2-\boldsymbol{q}^2{)}(\beta-1)}\right)^\frac{1}{2}\left(\frac{\beta-2}{\beta}\right),
}
\end{equation}
$C_2$ being an integration constant with the dimensions of a \textit{length}, the scalar $q_{\mu}x^{\mu}= q_{0}\ ct-q_{x}\ x- q_{y}\  y - q_{z}\  z$ being defined with four integration constants $q_\mu$.\\ 
Substituting the solution (\ref{tracer}) of the trace equation (\ref{ees3}) into the remaining equations (\ref{ees2}), and by taking into account (\ref{wfl}), we obtain weak-field limit of field equations for the physical unknowns $h_{\mu\nu}$ and $h$ in the de Donder gauge,
\begin{equation}
\label{h1}
\eqalign{
&\Box h^{(1)}_{\mu\nu}=C_{\mu\nu}{\xi}^{\frac{2}{\beta-2}},\\
&\Box h^{(1)}=C{\xi}^{\frac{2}{\beta-2}}
}
\end{equation}
which is straightforward integrated as
\begin{equation}
\eqalign{
\label{eq:34}
&h^{(1)}_{\mu\nu}=\Omega_{\mu\nu}\xi^{\frac{2(\beta-1)}{\beta-2}},\\
&h^{(1)}=\Omega \xi^{\frac{2(\beta-1)}{\beta-2}},
}
\end{equation}
where
we have defined the objects $C_{\mu\nu}$ and $\Omega_{\mu\nu}$ as  
\begin{equation}
C_{\mu\nu}\equiv-\eta_{\mu\nu}\frac{1}{3}\delta^{\frac{2}{\beta-2}}-2{q_{\mu}} {q_{\nu}}\gamma\beta{\delta^{\frac{2(\beta-1)}{\beta-2}}}\frac{2(\beta-1)\beta}{(\beta-2)^2},
\end{equation}
\begin{equation}
\label{eq:35}
\begin{array}{ll}
\fl \Omega_{\mu\nu}\equiv - \left(\frac{1}{\gamma}\right)^{\frac{1}{\beta-2}}& \left( \frac{(\beta-2)^2}{6(\beta-1)(q_{0}^{2} - \boldsymbol{q}^{2})}\right)^{\frac{\beta-1}{\beta-2}} \left(\frac{1}{\beta}\right)^{\frac{\beta}{\beta-2}} \eta_{\mu\nu} + \\  \\
&- \left(\frac{1}{2 \gamma}\right)^{\frac{1}{\beta-2}} \left(q_{0}^{2}-\boldsymbol{q}^{2} \right)^{\frac{3-2\beta}{\beta-2}} \left( \frac{1}{\beta}\right)^{\frac{\beta}{\beta-2}}  \left( \frac{(\beta-2)^2}{3(\beta-1)}\right)^{\frac{\beta-1}{\beta-2}}  q_{\mu}q_{\nu},
\end{array}
\end{equation}
and the contractions $C\equiv \eta^{\mu\nu}C_{\mu\nu}$, $\Omega\equiv \eta^{\mu\nu}\Omega_{\mu\nu}$.

The complete expression of $h_{\mu\nu}$ and $h$, defined in (\ref{harm}), is obtained by considering that $h^{(0)}_{\mu\nu}$ and $h^{(0)}$ obey, by construction, the field equations
\begin{equation}
\eqalign{
&\Box h^{(0)}_{\mu\nu}=0,\\
&\Box h^{(0)}=0.
}
\end{equation}

\paragraph{Discussion of the solution of the modified Einstein equations}
We have solved our field equations by first solving the trace equations, and then substituting it in the remaining equations. In this paragraph, we discuss the features of the solutions of the field equations, from the points of view of their physical interpretation and their mathematical properties.

It is possible to analyze the physical meaning of the trace equation as far as the presence of a scalar degree of freedom in modified-gravity theories is concerned. In fact, (\ref{ees3}) rewrites as
\begin{equation}\label{t1}
\Box\Phi=3\gamma\beta \Phi^{\frac{1}{\beta-1}}
\end{equation}
with $R\equiv\Phi^{\frac{1}{\beta-1}}$, as suggested by the presence of a scalar degree of freedom. Equation (\ref{t1}) is an inhomogeneous wave equation for the field $\Phi$. As a wave equation, we are suggested to look for a solution of the kind $\Phi(t,x,y,z,)\equiv\Phi(\xi)$, with $\xi$ defined in the above. Within this point of view, we then interpret its solution, according to its Fourier transform, as a linear superposition of non-linear functions of the wave-packets given by the massive modes.

We can infer that a function $f(\mathbf{R})$ which is not analytical in the neighborhood of $R\sim0$ implies a scalar degree of freedom ruled by a potential which is not analytical in the neighborhood of the value $R\sim0$ characterizing the weak-field limit.

So far, the solution of the trace equation (\ref{ees3}) is not in principle unique. Nevertheless, we see that, for the choice $R\equiv R(\xi)$, (\ref{ees3}) rewrites as an Emden-Fowler equation for the variable $w=R^{\beta-1}$, i.e. \cite{libro1} :
\begin{equation}\label{t2}
(q_0^2-\textbf{q}^2)\frac{d^2}{d\xi^2}w=3\gamma\beta w^{\frac{1}{\beta-1}}.
\end{equation} 
Emden-Fowler equations admit the general power-law solution (\ref{tracer}),
and, for certain values of the parameters, they may also admit a particular (parametric) solution. For the functional dependence of (\ref{t2}) on the parameters, no particular (parametric) solution is known to exist, and (\ref{tracer}) is therefore unique (\cite{libro1}, \cite{pz}).\\
We can also go the other way round and consider that the trace equation (\ref{ees3}) can be restated as a function of the unknown $w=R^{\beta-1}$,
for which \cite{libro1} the ``travelling wave solution in implicit form'' reads 
\begin{equation}
\label{implicit}
\fl \int{\left[C_{1}+\frac{2}{{{q_{0}}^2}-{\boldsymbol{q}^2}} \int{\frac{1}{3\gamma \beta}{w^{\frac{1}   {\beta-1}}}dw}\right]^{-\frac{1}{2}}dw}=q_{0} ct-q_x x-q_y y+ q_z z +  C_2
 \end{equation}
For a generic choice of the integration constants $C_1$ and $C_2$, (\ref{implicit}) is formally solved as
\begin{equation}
\label{eq:9}
\begin{array}{ll}
\fl \int{\frac{1}{\left[ 1+\frac{2(\beta-1)}{3\gamma {\beta^2} C_{1}({q_{0}^2}-{\boldsymbol{q}^2})}w^{\frac{\beta}{\beta-1}}\right]^{\frac{1}{2}}}dw}= \\ \\ = w\ {}_2 F_1 \left(
 \frac{1}{2}, \ \ \frac{\beta-1}{\beta}; \ \ \frac{2\beta-1}{\beta};\ \ -{\frac{2(\beta-1)}{3\gamma {\beta^2} C_{1}({q_{0}^2}-{\boldsymbol{q}^2})}}w^{\frac{\beta}{\beta-1}}\right),
\end{array}
\end{equation}
where ${}_2 F_1$ denotes the Gauss hypergeometric function \cite{libro1} \cite{libro2}.
For some special values of its arguments, the ${}_2 F_1$ function can be expressed in terms of elementary functions, but it is easily checked that, due to the range of the $\beta $ values ($2 < \beta < 3$), in our case it is not \cite{libro2}. Nevertheless, we notice that, by choosing $C_1\equiv0$, (\ref{implicit}) admits the explicit solution (\ref{tracer}).

\section{Interaction of modified gravitational waves with test particles\label{section5}}
We now use the geodesic deviation equation to study how the separation vector between two particles, $A$ and $B$, changes because of the presence of a modified gravitational wave. In general, we have:
\begin{equation}
\label{eq:GDA}
\frac{D^2}{D \tau^2}\delta x^{\mu}={R^{\mu}}_{\alpha \beta \gamma} U^{\beta} U^{\alpha} \delta x^\gamma,
\end{equation}
where $\delta x^{\mu}$ represents the separation vector between the two considered particles.\\
A convenient coordinate system for analyzing the previous equation is the proper reference frame of one of the two particles, say the particle $A$. This frame has spatial origin $x^{\widehat{j}}=0$ attached to A's geodesic and time coordinate equal to A's proper time, hence $x^{\widehat{0}}=\tau$ on the geodesic given by $x^{\widehat{j}}=0$; in addition, we assume that this frame is non-rotating  (see \cite{libro2}). The coordinate system we have just defined is a local Lorentz frame all along A's geodesic, so that:
\begin{equation}
\label{eq:LLF}
ds^2=d x^{{\widehat{0}}^2}- \delta_{\ {\widehat{i}}\ {\widehat{k}}} d x^{{\widehat{i}}} d x^{{\widehat{j}}}+ \mathcal{O}(| x^{{\widehat{j}}}|^2)d x^{{\widehat{\beta}}} d x^{{\widehat{\alpha}}}.
\end{equation}
Because of our choice of the reference frame, we also have $\delta x^{j}={x^{j}}_{B}$.\\
Moreover, at $x^{\widehat{j}}=0$, $ {\Gamma^{\widehat{\mu}}}_{{\widehat{\alpha}}\ {\widehat{\beta}}}$ vanish and so does $d {\Gamma^{\widehat{\mu}}}_{{\widehat{\alpha}}\ {\widehat{\beta}}} / d \tau$, thus, if we evaluate (\ref{eq:GDA}) along this geodesic, we find:
\begin{equation}
\label{eq:GDB}
\frac{d^2 x^{\widehat{j}}_{B} }{d \tau^2}= {R^{{\widehat{j}}}}_{{\widehat{0}}\ {\widehat{0}}\ {\widehat{k}}}\  x^{{\widehat{k}}}_{B}.
\end{equation}
 The curvature tensor is gauge invariant in the linearized theory, so we can write it by making use of the perturbation to the flat metric, given by ${h}_{\mu \nu}=h^{(0)TT}_{\mu \nu}+h^{(1)}_{\mu \nu}$, with $h^{(0)TT}_{\mu \nu}$ and $h^{(1)}_{\mu \nu}$ defined in the above. As a consequence, we obtain:
\begin{equation}
\label{eq:belR}
{R^{{\widehat{j}}}}_{{\widehat{0}}\ {\widehat{0}}\ {\widehat{k}}}=\frac{1}{2}\eta^{{\widehat{i}}{\widehat{j}}}\left(h^{(0)TT}_{\ {\widehat{i}}{\widehat{k}},{\widehat{0}}{\widehat{0}}}+h^{(1)}_{\ {\widehat{i}}{\widehat{k}},{\widehat{0}}{\widehat{0}}}+h^{(1)}_{\ {\widehat{0}}{\widehat{0}},{\widehat{i}}{\widehat{k}}}-h^{(1)}_{\ {\widehat{i}}{\widehat{0}},{\widehat{0}}{\widehat{k}}}-h^{(1)}_{\ {\widehat{0}}{\widehat{k}},{\widehat{i}}{\widehat{0}}}\right).
\end{equation}
In order to simplify the notation, we now drop all the symbols `` $\widehat{}$ ", which appear in (\ref{eq:GDB}) and (\ref{eq:belR}). Eventually, we get:
\begin{equation}
\label{eq:egde1}
\frac{d^2 x^{j}_{B}}{ d {\tau}^2}= \frac{1}{2}\eta^{ij}\left(h^{(0)TT}_{ik,00}+h^{(1)}_{ik,00}+h^{(1)}_{00,ik}-h^{(1)}_{i0,0k}-h^{(1)}_{0k,i0} \right)x^{{k}}_{B}.
\end{equation}
A key observation for solving (\ref{eq:egde1}) arise from the fact that we are treating a weak gravitational field, so we can also apply the slow motion approximation for the particles, so that:
\begin{equation}\label{uzero}
U^0 \sim 1 \ \ \ \ \ U^i \sim 0, \ \ \ \ i=1,2,3.
\end{equation}
Because of the relation (\ref{uzero}), we can make the following assumption:
\begin{equation}\label{appr}
x^0 =\tau,\ \ \ \ \ x^i (\tau)= \Delta^i,
\end{equation}
where the $\Delta^i$'s are constant quantities. The scalar $\xi$ rewrites $\xi=\left( q_0 c t + \Sigma \right)$, where we have defined the constant $\Sigma$ as $\Sigma\equiv\left(C_2 -q_x \Delta^1- q_y \Delta^2- q_z \Delta^3 \right)$.\\
Furthermore, we assume that, initially, the particles are at rest relative to each other, so that $
x^{{j}}_{B} \equiv x^{{j}}_{B (0)}$, and we consider the particle relative position displacement, induced by the perturbation ${h}_{\mu\nu}$, as a small perturbation with respect to the initial position $x^{{j}}_{B (0)}$. Due to this fact, we can set:
\begin{equation}
\label{eq:dv}
x^{{j}}_{B}=x^{{j}}_{B (0)}+x^{{j}}_{B(1)}(\tau),
\end{equation}
where $x^{{j}}_{B (0)}$ is a constant vector, and the index $j$ refers to spatial coordinates only. \\
Collecting all the ingredients together, solving (\ref{eq:egde1}), (\ref{eq:dv}) rewrites
\begin{equation}\label{ll}
\begin{array}{ll}
\fl x^{j}_{B}=x^{j}_{B(0)}+\frac{1}{2}\eta^{ji}h^{(0)TT}_{ik}x^{k}_{B(0)}&+\\ \\&+\frac{1}{2}\eta^{ji} \left[\Omega_{00}\frac{q_i q_k}{{q_{0}^2}}+\Omega_{ik}-\Omega_{i0}\frac{q_k}{q_{0}}-\Omega_{0k}\frac{q_i}{q_{0}}\right] \xi^{\frac{2(\beta-1)}{\beta-2}}x^{k}_{B(0)}.
\end{array}
\end{equation}
where: ${{i}}, {{j}}, {{k}}=1,2,3$, thus $q_i=-q^i=(-q_x, -q_y, -q_z)$ and $q_0=q^0$.

\paragraph{Geodesic displacements}
To study the features of the modified gravitational waves in our model, we choose particular initial conditions for the wave, such that the effects of the modified gravitational wave are easily pointed out. Furthermore, we will choose the most appropriate numerical values of the parameters, such that the constraints on the modified part of the gravitational wave are the strictest. We will also illustrate the effects of the extra polarization modes of the gravitational wave.\\
To do so, in particular, we assume the $h^{(0)TT}_{\mu\nu}$ part as propagating along the $x$ direction, in the $A_{+}$ polarization, and the $h^{(1)}_{\mu\nu}$ part as propagating along the $y$ direction. The relative displacement of the particle B under the influence of the gravitational wave, $(x^i_B-x^i_{B(0)})/x^i_{B(0)}$ are given by:
\numparts
\begin{equation}
\label{eq:srdm1b}
\frac{x^{{1}}_{B}-x^{{1}}_{B (0)}}{x^{{1}}_{B(0)}}=-\frac{1}{2}a\xi^{\frac{2(\beta-1)}{\beta-2}},
\end{equation}
\begin{equation}
\label{eq:srdm2b}
\frac{x^{{2}}_{B}-x^{{2}}_{B (0)}}{x^{{2}}_{B (0)}}=+\left[-\frac{h^{(0)TT}_{22}}{2}\right] -\frac{1}{2}\left[\frac{{q_0}^2-{q_2}^2}{{q_0}^2}a\right]\xi^{\frac{2(\beta-1)}{\beta-2}},
\end{equation}
\begin{equation}
\label{eq:srdm3b}
\frac{x^{{3}}_{B}-x^{{3}}_{B (0)}}{x^{{3}}_{B(0)}}=+\left[-\frac{h^{(0)TT}_{33}}{2}\right] -\frac{1}{2}a\xi^{\frac{2(\beta-1)}{\beta-2}},
\end{equation}
\endnumparts
where the terms containing $\xi$ come from $h^{(1)}_{\mu\nu}$.\\
From the previous relations, it is possible to check that we are allowed to define the velocity $v$ of the GW modified part $h^{(1)}_{\mu\nu}$ as $v^2=c^2q_0^2/q_y^2$. In order to obtain a numerical estimation for $v$, we now assume to reveal GWs today, i.e. at $t=t_u\equiv 14 \times 10^9$ years, and at $y=x^{2}=0$. Moreover, we note that $\xi=q_{\mu}x^{\mu}+C_2$, where $C_2$ is an arbitrary constant, then, we also make the assumption that, for our choices, we can neglect $C_2$, or set $C_2=0$.\\
Since there are not direct evidences of the gravitational waves, like, for instance, observations by means of ground-based interferometers, we can also assume that the modified part of the GWs in our $f(R)$ gravity cannot produce relative displacements larger than those produced by the ordinary gravitational waves.
If we now consider $A_{+}\sim5\cdot10^{-23}$ for the ordinary part $h^{(0)TT}_{\mu\nu}$ \footnote{According to (9) of \cite{PhysAstrophCosmGWs}, where we are using $G=6.67 \times 10^{11} m^3 kg^{-1} s^{-2}$ and $c=3 \times 10^8 m/s$, it is possible to obtain an amplitude of the standard GWs of about $10^{-23}$, if we use the following approximated values for a binary star system: Object masses $= M_{\odot}$, Orbital radius $= 20 \times 10^{3}m$, Orbital frequency $= 400 Hz$, Source distance $= 100Mpc$.}, we can associate with this term relative displacements of the order of $10^{-23}$. Thus, setting $\beta\equiv2+n$, we obtain the following expressions $v_2$ from (\ref{eq:srdm2b}) and $v_{1,3}$ from (\ref{eq:srdm1b}) and (\ref{eq:srdm3b}) :
\numparts
\begin{equation}
\label{eq:ll4}
{v_{2}}^2 = \frac{c^2}{1+\frac{c^2}{\frac{10^{-23n}}{c^{2n}\left({\frac{1}{2}}\right)^{n}\left(\frac{n^2}{6+6n}\right)^{1+n}\left(\frac{1}{2+n}\right)^{2+n}\left(t/L_{\gamma}\right)^{2(1+n)}}}},
\end{equation}
\begin{equation}
\label{eq:lll4}
{v_{1,3}}^2 = \frac{c^2}{1+\frac{c^2}{\frac{10^{-\left(23\frac{n}{n+1}\right)}}{\left({\frac{1}{2}}\right)^{\frac{n}{n+1}}\left(\frac{n^2}{6+6n}\right)\left(\frac{1}{2+n}\right)^{\frac{2+n}{n+1}}\left(t/L_{\gamma}\right)^{2}}}}.
\end{equation}
\endnumparts
 For the picture considered above, after fixing an allowed value for $L_{\gamma}$, (\ref{eq:ll4}) and (\ref{eq:lll4}) depend only on $\beta$, thus it is possible to determine numerical estimations for both ${v_{2}}$ and ${v_{1,3}}$. Hereafter, we will take into account an intermediate value for the characteristic length scale of our model, i.e. $L_{\gamma}=38pc$. It is worth noting that,
in the limit $\beta\rightarrow2$, we obtain that both $v_2$ and $v_{1,3}$ tend to $c$, while, for $\beta \rightarrow 3$, we get $v_{2} \sim 6.197 \times 10^{-21} km/s$ and ${v_{1,3}} \sim 4.3 \times 10^{-8} km/s$. For the discussion in the abaove, (\ref{eq:ll4}) and (\ref{eq:lll4}) give us the maximum value the velocity $v$ can have, so that we do not obtain relative displacements, due to ${h^{(1)}_{\mu\nu}}$, larger than those expected from the ordinary gravitational waves.  Because of the different dependence on $n$ and because of the presence of the factor $c^{2n}$, it is possible to check that (\ref{eq:ll4}) requires a smaller value of the velocity with respect to (\ref{eq:lll4}). As a consequence, we have to assume as a good estimation for $v$, the values obtained from (\ref{eq:ll4}), indeed it imposes more restrictive constraints.

\begin{figure}[ht!]
\centering
\includegraphics[width=0.95\textwidth, height=0.42\textheight]{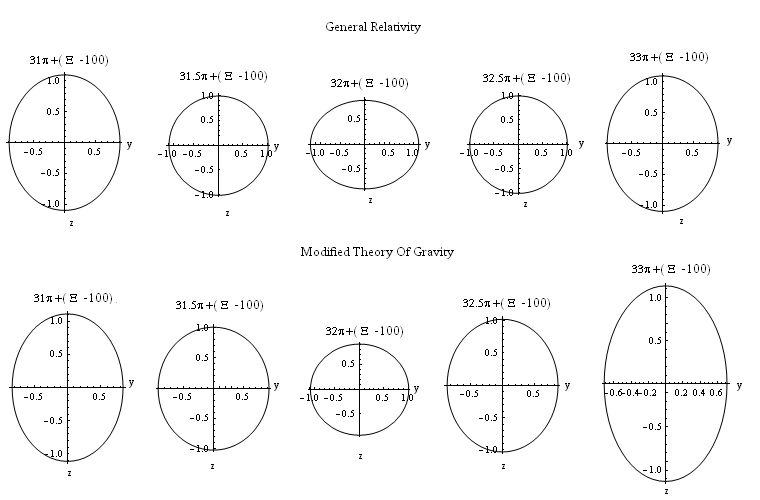}
\caption{\label{fig:qpp} \small{Left column: polarization ellipses in General Relativity. Right column: polarization ellipses for $f(R)=R+\gamma R^{\beta}$ gravity theory when $\beta=2.0000001$ and $L_{\gamma}=38pc$. For the sake of clearness, we note that we have used the same argument $31\pi+(\Xi-100)\leq p=\leq33 \pi+(\Xi-100)$ for both the two terms of the modified GWs and GWs in General Relativity.}}
\end{figure}

\begin{figure}[ht!]
\centering
\includegraphics[width=0.98\textwidth, height=0.48\textheight]{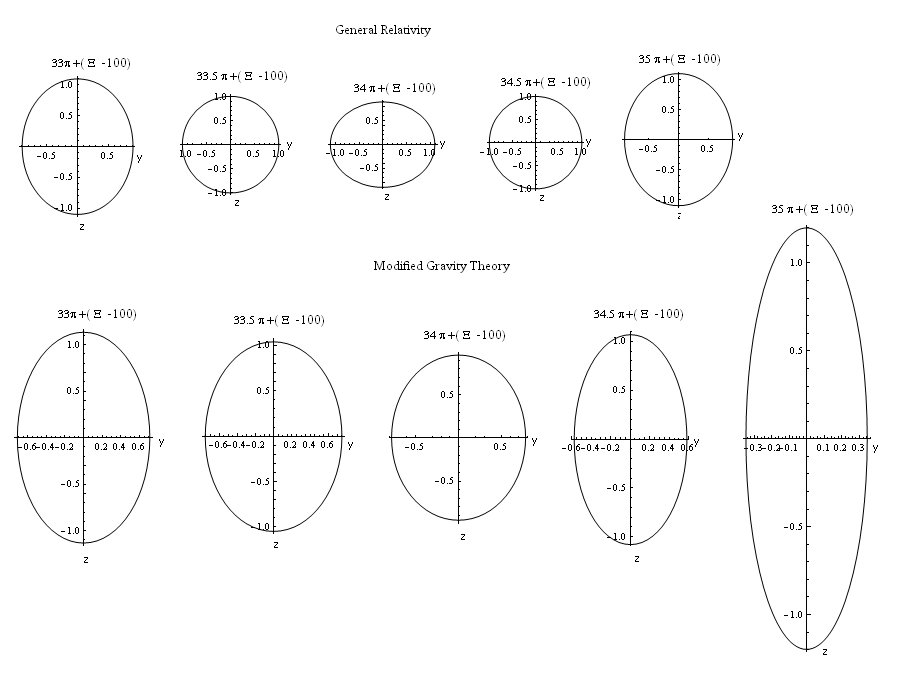}
\caption{\label{fig:qpg}\small{Left column: polarization ellipses in General Relativity. Right column: polarization ellipses for $f(R)=R+\gamma R^{\beta}$ gravity theory when $\beta=2.0000001$ and $L_{\gamma}=38pc$. For the sake of clearness, we note that we have used the same argument $33 \pi+(\Xi-100)\leq p=p_{GR}\leq35 \pi+(\Xi-100)$ for both the two terms of the modified GWs and GWs in General Relativity. }}
\end{figure}

\paragraph{Modified polarization} We now want to show that the modified gravitational waves, described by the two terms $h^{(0)TT}_{\mu\nu}$ and ${h^{(1)}_{\mu\nu}}$, change the polarizations of the standard gravitational waves. Here, we will keep on assuming that the term $h^{(0)TT}_{\mu\nu}$ describes an ordinary gravitational wave propagating along the $x$-axis with plus polarization only and the modified part of the gravitational waves, indicated by $h^{(1)}_{\mu\nu}$, is charactered through the quadrivector $q^{\mu}=(q^0,0,q^2,0)$. With these assumptions, (\ref{eq:srdm1b}), (\ref{eq:srdm2b}) and (\ref{eq:srdm3b}) are still valid; furthermore, if we consider the arbitrary constant $C_2$ as negligible or set it to zero, we have:
\begin{equation}
\label{eq:perxi}
 \xi=cq_0\left(t-y/v\right).
\end{equation}
By making use of the relations  (\ref{eq:srdm1b}), (\ref{eq:srdm2b}), (\ref{eq:srdm3b}) and (\ref{eq:perxi}), it is possible to verify that we can define the argument of the GW modified part as the dimensionless quantity $(c/L_\gamma)(t-y/v)$. We now observe that, in our calculation, we are going to use the previous velocity estimations. As a consequence, we point out that we estimated the velocities of the GW modified part, by imposing that, at $y=0$ and $t_u=14\times 10^{9}$years, the relative displacements due to the term $h^{(1)}_{\mu\nu}$ were of the same order of the relative displacements associated with $h^{(0)TT}_{\mu\nu}$. Since we want to show the effects introduced by the term $h^{(1)}_{\mu\nu}$ on the ordinary gravitational wave polarizations, we need to consider only those situations where the observable effects due to the term $h^{(1)}_{\mu\nu}$ are at least comparable with the observable effects due to the term $h^{(0)TT}_{\mu\nu}$. Hence, we will take as reference value for the argument $(c/L_\gamma)(t-y/v)$ of the GW modified part the one we have used when deriving the velocity estimations, that is:\begin{equation}
\Xi=\frac{c}{L_{\gamma}}\left(t-\frac{y}{v}\right)= \frac{ct_u}{L_{\gamma}}=1.16184 \times 10^8,
\end{equation}
where we keep on using $L_{\gamma}=38pc$.
As far as the relative displacements due to the term $h^{(0)TT}_{\mu\nu}$ are concerned, for the particular conditions we are considering, we will use the following definition:
\begin{equation}
\label{eq:forst}
\frac{h^{(0)TT}_{22}}{2}=-\frac{h^{(0)TT}_{33}}{2}=h_{+}{\cos}\left(\omega\left(t-\frac{x}{c}\right)\right),
\end{equation}
with $h_{+}=10^{-23}$.
To visualize the effects introduced by the term $h^{(1)}_{\mu\nu}$ on the GW polarization, we have taken into account a system of particles initially arranged on a circumference, on the plane $x=0$.
Here, we insert the plots for this case (see figure (\ref{fig:qpp}) and figure (\ref{fig:qpg})), where we have amplified the effects due to the modified GWs, for the sake of an effective comparison.

In figures (\ref{fig:qpp}) and (\ref{fig:qpg}), as far as the argument of the modified GW  is concerned, we have adopted the same values for both $h^{(0)TT}_{\mu\nu}$ and $h^{(1)}_{\mu\nu}$.\\ 
Our results tell us that, in the case of $f(\bf R)=\bf R+\gamma \bf R^{\beta}$ theory, we obtain polarization ellipses which change with time, similarly to the General Relativity case. However, differently from General Relativity, in our $f(\bf R)$ gravity theory, the polarization ellipses keep changing and do not assume periodically the same shape. We also note that the larger the argument values, the greater the effects of the GW modified part. This last result comes from the fact that the amplitude of the modified GW term associated with  $h^{(1)}_{\mu\nu}$ increases when the argument increases.

Moreover, from (\ref{eq:srdm1b})-(\ref{eq:srdm3b}), we can see that, in general,  modified GWs have effects along all the three spatial directions, then GWs, in $f(\bf R)=\bf R+\gamma \bf R^{\beta}$ gravity, are not transverse.

\section{Concluding remarks}
In this paper, we have analyzed the features of the model $f(\mathbf{R})=\mathbf{R}+\gamma \mathbf{R}^\beta$, whose Taylor expansion in the weak-field limit $R\sim0$ does not hold.\\
In Section \ref{section2}, we have pointed out the main features of modified $f(R)$ gravity, focusing our attention on the characteristics of the weak-field limit, as far as analytical models as well as non-analytical ones are concerned.\\
In Section \ref{section3}, we have reviewed in some detail the properties of the non-analytical model $f(\mathbf{R})=\mathbf{R}+\gamma \mathbf{R}^\beta$, which have already been investigated in previous works, i.e. \cite{Lecian:2008vc} for the weak-field limit in the spherically-symmetric static case, and \cite{Capozziello:2011yr} for the cosmological implementation.\\
In Section \ref{section4}, we have found the analytical solution of the weak-field limit of field equations as far as the presence of gravitational waves is concerned. To do so, we have briefly recalled the properties of some useful gauges in GR, and their effects on our modified field equations. After this, we have summarized the main results found in the case of an $f(\mathbf{R})$ model whose Taylor expansion in the weak-field limit around the value $R\sim0$ holds. The explicit solution of the field equations accounting for gravitational waves has then been given, and its features have been investigated and commented as far as their physical interpretation is concerned, as well as its geometrical structure and its mathematical well-posed-ness.\\
In section \ref{section5}, we have studied the interaction of the gravitational waves described by the $f(\mathbf{R})=\mathbf{R}+\gamma \mathbf{R}^\beta$ model with test particles. More precisely, we have solved the geodesic deviation equation and have imposed some constraints on the velocity of the modified gravitational wave. Furthermore, we have discussed the polarizations which are present in this model.\\
Thus, we have shown how the particular modification
of General Relativity we addressed here is able to
provide a very peculiar trace of its presence by
means of the emergence of a non-linear and
non-negligible correction to the linear
theory of the gravitational waves propagation.
Such a specific morphology of the modified
spacetime ripples, more than in their non-transverse and
non-traceless behaviour
(present in other modified gravitational wave
paradigms)
, consists in the details
of their polarization and propagation features.
In fact, our study of the polarization tensor,
together with the increasing power-law behaviour
of the modified term, induce a well-determined
picture of the polarization ellipses, which seems
the very smoking gun of the modified
power-law Lagrangian we added to the standar Einstein-Hilbert
shape. This striking morphology, well-grounded
on the amplification that any signal of this
sort acquires in a given space point, make
the issue presented above of some impact also
for the expectation of the future detection of
gravitational signals, even because the
proposed non-linear spacetime ripples are
specifically traced in the stochastic sea of all
the other astrophysical signal, at the lowest order,
properly accounted
by General Relativity.

\ack
The authors kindly thank Riccardo Benini for his fruitful
suggestions.
DF gratefully thanks  Eric Chassande-Mottin for having brought \cite{PhysAstrophCosmGWs} to her attention. OML warmly thanks the Max Planck Institute - Albert Einstein Institute for great hospitality during the final stages of this work and Cecilia Chirenti for remarking the possibility to use \cite{eloisa} for a comparison with modified theories of gravity. The work of DF and GM was partially developed within the framework of the CGW Collaboration (http:// www.cgwcollaboration.it).

\end{document}